\documentclass{emulateapj}
\usepackage{times,mathptmx}

\usepackage{graphicx}

\newcommand{\ha}{\mbox{H$\alpha$}}
\newcommand{\hb}{\mbox{H$\beta$}}

\newcommand{\hii}{\hbox{H {\sc ii}}}
\newcommand{\oii}{\hbox{[O {\sc ii}]}}
\newcommand{\oiii}{\hbox{[O {\sc iii}]}}
\newcommand{\nii}{\hbox{[N {\sc ii}]}}

\newcommand{\etal}{et al.\,}

\newcommand{\ergsec}{\hbox{erg sec$^{-1}$}}
\newcommand{\microjy}{\hbox{$\mu$Jy}}
\newcommand{\stwofour}{\hbox{$S_{24}$}}

\shortauthors{Weiner et al.}
\shorttitle{Extinction and Star Formation Tracers}

\slugcomment{Accepted by ApJL}

\begin{document}

\title{AEGIS: Extinction and Star Formation Tracers from Line Emission}

\author{Benjamin J. Weiner\altaffilmark{1,2},
Casey Papovich\altaffilmark{2,8},
K. Bundy\altaffilmark{3},
C.J. Conselice\altaffilmark{4},
M.C. Cooper\altaffilmark{5},
R.S. Ellis\altaffilmark{3},
R.J. Ivison\altaffilmark{6},
K.G. Noeske\altaffilmark{7},
A.C. Phillips\altaffilmark{7},
Renbin Yan\altaffilmark{5}
}

\altaffiltext{1}{Dept. of Astronomy, University
of Maryland, College Park, MD 20742}
\altaffiltext{2}{Steward Observatory, University of Arizona, Tucson, AZ 85721, {\tt bjw@as.arizona.edu}}
\altaffiltext{3}{California Institute of Technology, Pasadena, CA 91101}
\altaffiltext{4}{University of Nottingham, NG7 2RD, UK}
\altaffiltext{5}{Dept. of Astronomy, University of California, Berkeley, 
Berkeley, CA 94720}
\altaffiltext{6}{Institute for Astronomy, University of Edinburgh,
                 Blackford Hill, Edinburgh EH9\,3HJ, UK}
\altaffiltext{7}{UCO/Lick Observatory, University of California, Santa Cruz, 
Santa Cruz, CA 95064}
\altaffiltext{8}{Spitzer Fellow}

\begin{abstract}

Strong nebular emission lines are a sensitive probe of star formation
and extinction in galaxies, and the \oii\ line detects star forming 
populations out to $z>1$.  However, star formation rates
from emission lines depend on calibration of extinction and
the $\oii/\ha$ line ratio, and separating star formation from 
AGN emission.  We use calibrated line luminosities from the DEEP2
survey and Palomar $K$ magnitudes to show that the behavior of emission
line ratios depends on galaxy magnitude and color.  For galaxies
on the blue side of the color bimodality, the vast majority show
emission signatures of star formation, and there are strong
correlations of extinction and $\oii/\ha$ with restframe $H$ magnitude.
The conversion of \oii\ to extinction-corrected $\ha$ and thus to
star formation rate has a significant slope with $M_H$, 0.23 dex/mag.
Red galaxies with emission lines have a much higher scatter in their
line ratios, and more than half show AGN signatures.
We use 24 $\mu$m fluxes from Spitzer/MIPS to demonstrate the
differing populations probed by nebular emission and by mid-IR luminosity.
Although extinction is correlated with luminosity, 98\% of
IR-luminous galaxies at $z \sim 1$ are still detected in
the \oii\ line.  Mid-IR detected galaxies are mostly bright and
intermediate color, while fainter, bluer galaxies with high \oii\ 
luminosity are rarely detected at 24 $\mu$m.

\end{abstract}


\section{Introduction}

Strong nebular emission lines such as \ha, \hb, and \oii\ 3727 \AA\
arise in \hii\ regions and are
indicators of the star formation rate (SFR).  These lines, especially
\oii, enable the measurement of star-forming galaxies out to $z>1$ in
optical spectroscopy 
(e.g.\ Brinchmann \& Ellis 2000; Bauer \etal\ 2005).
However, SFRs from nebular lines are sensitive
to the calibration models, to extinction, and to AGN/non-star-formation 
emission (see e.g.\ Kennicutt 1998).  Other probes 
such as radio and mid- to far-IR luminosity are less affected by 
extinction, but test different timescales of star formation and 
generally do not reach SFRs as low as emission lines and UV
luminosity do (e.g.\ Bell \etal\ 2005).

Redshift surveys with multiwavelength data offer several ways
to measure the evolution of the SFR in galaxies to $z>1$.
With data from the UV, optical spectra, mid-IR and radio, we 
can compare multiple SFR indicators and
their calibrations, reliability, and biases. 
This paper takes a first step by examining 
optical emission lines as a function of galaxy magnitude
and color to measure extinction, $\oii/\ha$ calibration,
and AGN fraction, and by comparing to mid-IR data to test
for the prevalence and detectability of obscured star
formation.

\section{Sample and Data Extraction}

Our sample is drawn from the All-Wavelength Extended Groth Strip 
International Survey (AEGIS) (Davis \etal\ 2006).
We limit the sample to 9505 galaxies with reliable redshifts from
DEEP2 spectra.
6870 of these have $K$ magnitudes from Palomar/WIRC 
observations, which reach $K_s \sim 23$ (AB) at $5\sigma$.
5421 of these are in an area covered by both DEEP2 and Spitzer/MIPS 
observations, and 1645 have 24 $\mu$m detections, to a
flux limit of $\stwofour \sim 83$ \microjy.
The galaxies not detected in $K_s$ are generally the
faintest and bluest of the DEEP2 sample (Bundy \etal\ 2006).  
Restframe $M_B$ and $U-B$ (AB, $\Lambda$CDM) are computed from $BRI$ 
by SED fitting (Weiner \etal\ 2005; Willmer \etal\ 2006),
and rest $M_H$ from $I$ and $K_s$, using model SEDs
from PEGASE (Fioc \& Rocca-Volmerange 1997).

We fit emission lines in DEEP2 spectra with a
non-linear least squares fit of a gaussian (or doublet for \oii) 
to data in a 15 \AA\ radius about
the line.  For a robust continuum we use the biweight of the data 
20-60 \AA\ from the line.  
We use the $R-I$ color,
$I$ magnitude, and our $K$-correction procedure with the SEDs 
of Kinney \etal\ (1996) to derive an AB magnitude in the 
continuum windows
and combine this with the line equivalent width for a flux-calibrated
line luminosity in \ergsec.  This procedure compensates for 
throughput and slit losses if the line/continuum ratio is constant 
between the light in the slit and the entire object, reasonable for 
DEEP2 galaxies, of which 95\% have $r_{eff}<0.95\arcsec$.

Corrections for underlying stellar Balmer absorption
depend on spectral resolution and fitting method (e.g. 5 \AA\ in 
Kennicutt 1992, 1-1.5 \AA\ in Kewley \etal\ 2002).  
Because stellar Balmer absorption lines are broad
and DEEP2 spectra are 1.4 \AA\ FWHM, we
effectively fit the emission as it sits in the absorption
trough (cf. Figure 2 of Choi \etal\ 2006).
Here, we do not apply a correction.
The median restframe EWs for blue galaxies that have both \ha\ and
\hb\ are 37 and 8 \AA\ respectively, on the corrected relation
of Kennicutt (1992).  

Because the DEEP2 spectral range is limited to about 6600-9200 \AA,
we measure line pairs such as \ha\ and \hb\ or \oii\ and
\hb\ only in narrow ranges of redshift, and cannot
measure $\oii/\ha$ directly.  We also measure line strengths 
in the Team Keck survey (TKRS) 
of the GOODS-N field (Wirth \etal\ 2004), which has DEIMOS
spectra with wider wavelength coverage, of a sample selected
similarly to DEEP2, and apply $K$-corrections
to the magnitudes of Capak \etal\ (2004).

\section{Behavior of extinction and star formation tracers}

\subsection{Emission line dependence on magnitude and color}

\begin{figure}[t]
\begin{center}
\includegraphics[width=3.5truein]{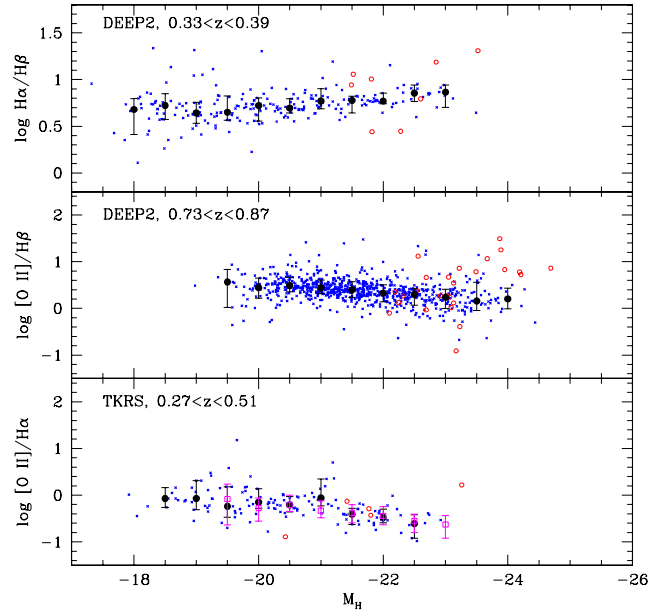}
\caption{Line luminosity ratios as function of $M_H$ (AB).
Blue galaxies with $U-B <1.10-0.032(M_B+21.5)$ are Xes and 
red galaxies are circles.  The large points are the
median and $\pm 34\%$ in magnitude bins for the blue galaxies only.
For $\oii/\ha$, the large circles are the median of the TKRS galaxies
and the squares are the trend inferred for DEEP2
by combining the $\hb/\ha$ and $\oii/\hb$ median trends.
Redshift ranges in each panel indicate the range containing
95\% of the galaxies with that line ratio.
}
\label{fig-deepmhlineratio}
\end{center}
\end{figure}

We begin by examining emission lines commonly
used to measure extinction and trace star formation.
Figure \ref{fig-deepmhlineratio} shows the emission line ratios
$\ha/\hb$ and $\oii/\hb$ in energy units as functions of 
restframe IR magnitude $M_H$.  Only ratios with error $<0.5$ dex are 
used, to remove poor fits due to lack of emission or sky contamination,
eliminating $\sim 40\%$ of red galaxies and $\sim 5\%$ of blue.  
The point type indicates whether the galaxy is on the blue or 
red side of the color bimodality, $U-B=1.10-0.032(M_B+21.5)$ 
(Willmer \etal\ 2006).
There is a distinct difference between blue and red galaxies:
the blue galaxies have well defined trends with $M_H$ while
the red galaxies have about twice as much scatter.  

Blue galaxies have a well-defined
distribution of $\ha/\hb$, with a
measurable but weak dependence on $H$ magnitude. 
The Balmer decrement $\ha/\hb$ is a measure of the extinction;
a linear fit to only the blue galaxies gives log $\ha/\hb$ = 0.74
at $M_H=-21$, with a slope of $-0.040 \pm 0.006$ dex/$H$ mag, so that
nebular extinction is greater in brighter galaxies. 
At $M_H=-21$, the reddening is $E(\hb-\ha) = 0.29$ dex and the
extinction of \ha\ is $A(\ha)=0.68$ dex, with a slope of --0.094 
dex/$H$ mag, assuming an unreddened $\ha/\hb=2.86$ and the extinction 
law of Cardelli, Clayton \& Mathis (1989) with $R_V=3.1$.
The RMS in $\ha/\hb$ about the fit is 0.15 dex, of which 0.1 dex
is intrinsic scatter beyond the errors.
The slope of the relation
is consistent with estimates in nearby and SDSS galaxies
(Jansen \etal\ 2001; Stasi{\'n}ska \etal\ 2004; Moustakas,
Kennicutt \& Tremonti 2006).

In the middle panel of Figure \ref{fig-deepmhlineratio} we show
the $\oii/\hb$ ratio.  This ratio is affected by metallicity,
excitation, and reddening, causing it to decline
with galaxy mass.  For the blue galaxies, a linear fit
yields log $\oii/\hb$ = 0.43 at $M_H=-21$ and slope 
$+0.094 \pm 0.008$ dex/mag,
with rms 0.23 dex; the intrinsic scatter is 0.17 dex.
Given the slope of $\ha/\hb$ with $M_H$,
45\% of the slope in $\oii/\hb$ is due to reddening.

It is very desirable to measure the $\oii/\ha$ ratio,
for purposes of testing SFR
calibrations (e.g. Kennicutt 1992; Hopkins et al 2003; 
but see Kewley \etal\ 2004).  The lower panel of 
Figure \ref{fig-deepmhlineratio} shows the $\oii/\ha$ ratio
for galaxies in the smaller TKRS survey.  The large black points
are the median and RMS of the TKRS galaxies; the large magenta
squares are the median trend inferred by combining the
DEEP2 $\ha/\hb$ and $\oii/\hb$ trends.  Despite the different
$z$-ranges, the inferred DEEP2 and 
real TKRS $\oii/\ha$ trends are very similar;
Moustakas \etal\ (2006) found that $z\sim 1$ galaxies are
consistent with a local relation.
It appears that the trends with $M_H$ are consistent because the 
line ratios' dependence on mass/luminosity is stronger than the 
evolution with $z$; the evolution in mean metallicity at fixed mass
to $z\sim 0.8$ is relatively small, compared to the overall 
trend with luminosity and to the scatter 
(Kobulnicky \& Kewley 2004).

In the AEGIS/DEEP2 data, for galaxies at $z \lesssim 1$, combining
$\ha/\hb$ and $\oii/\hb$ yields an inferred log $\oii/\ha = -0.32$
at $M_H=-21$ with a slope of 0.134 dex/mag and scatter $\sim 0.3$ dex.
A fit to the TKRS data yields log $\oii/\ha = -0.28$ at $M_H=-21$ with
a slope of $0.129 \pm 0.02$ dex/mag.
The slopes are mildly shallower than Jansen \etal\ (2001) and similar
to Moustakas \etal (2006) for $\oii/\ha$ versus $M_B$.
At $M_H=-21$, $F(\oii)/F(\ha) = 0.48$, similar to the sample of
Kennicutt (1992), but twice the ratio in the sample of Hopkins \etal (2003)
and 0.6 that implied by Gallagher \etal\ (1989).
Because the slope with $M_H$ is significant, the $\oii/\ha$ conversion 
has a tilt, reinforcing the point that SFR calibrations for
\oii\ can be systematically off if an inapplicable $\oii/\ha$ ratio
is used (Kewley \etal\ 2004).  
The slope of \oii\ observed/\ha\ extinction-corrected, 
$\oii_{obs}/\ha_{extcorr}$, is +0.23 dex/mag.
SFR measurements from \oii\ as a function of mass will be biased if 
these corrections for $\oii/\ha$ and extinction are omitted.

\subsection{Emission line discriminators of AGN}

\begin{figure}[t]
\begin{center}
\includegraphics[width=3.5truein]{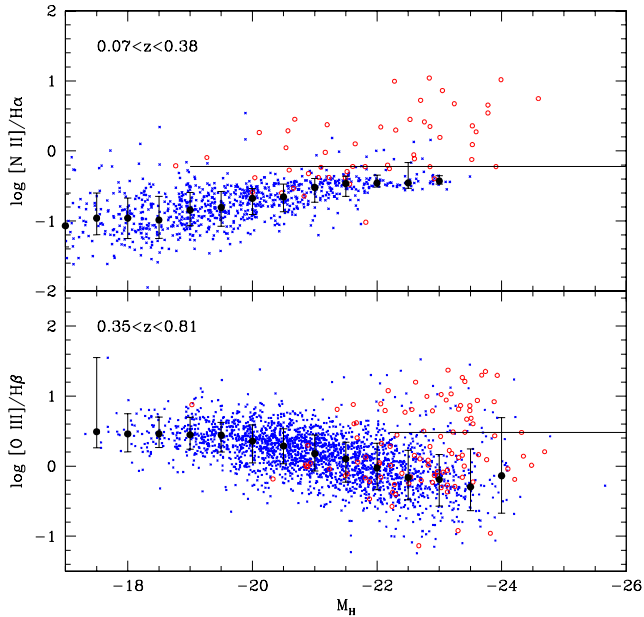}
\caption{
DEEP2 line ratios $\nii/\ha$ and $\oiii/\hb$ as a function of $M_H$ (AB).
Blue galaxies are Xes and red galaxies are circles.  The large points are
the median and RMS in magnitude bins for the blue galaxies only.
Horizontal lines indicate the AGN criterion $\nii/\ha>0.6$ and 
AGN/LINER division $\oiii/\hb=3$.
}
\label{fig-mh-bpt}
\end{center}
\end{figure}

A potential problem for many star formation indicators
is confusion with AGN light.  Star formation and
AGN/LINERs can be distinguished by emission 
line signatures.  Figure \ref{fig-mh-bpt}
plots two components of an AGN diagnostic diagram,
the $\nii/\ha$ and $\oiii/\hb$ ratios
(Baldwin, Phillips \& Terlevich 1981; Veilleux \& Osterbrock 1987).
These ratios are sensitive to metallicity and excitation but
not to reddening.


Blue galaxies follow a well-defined trend of increasing
$\nii/\ha$ with luminosity, reflecting the mass-metallicity
relation, even though $\nii/\ha$ is also sensitive to ionization
parameter (Kobulnicky \& Kewley 2004).  Some red galaxies 
fall on the blue star-forming galaxy track, but 55\% of red 
galaxies with emission and 1\% of blue galaxies 
are above the AGN criterion of
$\nii/\ha>0.6$, indicating higher ionization
(Veilleux \etal\ 1995; Kauffmann \etal\ 2003).

The lower panel of Figure \ref{fig-mh-bpt} shows that
in blue galaxies $\oiii/\hb$ depends on $M_H$, with a strong trend and 
a high RMS scatter of 0.3 dex about it.  Within the sample, the brighter
galaxies evolve, decreasing by 0.3 dex in $\Delta z = 0.4$.
This plot is reminiscent of a ``BPT'' diagnostic diagram (Baldwin 
\etal\ 1981) of $\oiii/\hb$ versus
$\nii/\ha$; here $M_H$ is on the x-axis instead of $\nii/\ha$.
Because $M_H$ is correlated with $\nii/\ha$,
the lower panel is a ``pseudo-BPT'' diagram.

In the BPT diagram, 
star-forming galaxies are on a track of declining $\oiii/\hb$
with increasing $\nii/\ha$, similar to the blue galaxies here,
due to metallicity and excitation trends with mass.
AGN and LINERs are on a spur at high $\oiii/\hb$ and high $\nii/\ha$;
the horizontal line indicates an AGN/LINER separation at $\oiii/\hb=3$
(e.g. Baldwin \etal\ 1981; Veilleux \etal\ 1995;
Kauffmann \etal\ 2003; Tremonti \etal\ 2004).
In the pseudo-BPT diagram, the separation between AGN and star-formers 
is not quite as good because we lose the information provided by
enhanced $\nii/\ha$.
Still, the vast majority of blue galaxies are on the star-forming
track, and 2-3\% of blue galaxies sit above it.  The red
galaxies detected in emission have a very high
scatter in $\oiii/\hb$. Some red galaxies overlap
the star-forming track, but about half of them are well 
above it on the AGN-like spur, where \oiii\ is enhanced by
higher ionization.

The pseudo-BPT diagram indicates that
line emission in the vast bulk of DEEP2 blue galaxies is from star
formation, while emission in red galaxies,
especially the brighter (and redder) red galaxies, is quite
frequently due to AGN.  Locally, AGN are preferentially found in 
massive galaxies with strong \oiii\ (Kauffmann \etal\ 2003) and
emission lines from red galaxies are generally due to AGN rather 
than star formation (Yan \etal\ 2006); X-ray selected AGN
are mostly red (Nandra \etal\ 2006).  There are some red 
galaxies on the star-forming tracks of Figure \ref{fig-mh-bpt}.
Morphologies
of red galaxies to $z\sim 1$ show that most are spheroidals, but 
$\sim 10\%$ are edge-on disks and $\sim 10\%$ are diffuse or 
irregular (Bell \etal\ 2004; Weiner \etal\ 2005).  Both of
these latter categories are likely dusty and star-forming.
Based on the $\nii/\ha$ ratio, $\sim 1/3$ of red galaxies with
emission are close to the star-forming track.

\subsection{Infrared tracers, extinction, and detection limits}

Extinction can have a strong effect on optical, UV, and emission-line
star formation indicators; locally extinction
is high in high-SFR galaxies, but with a high scatter
(e.g. Wang \& Heckman 1996; Heckman \etal\ 1998; Calzetti 2001;
Hopkins \etal\ 2001; Sullivan \etal\ 2001;
Afonso \etal\ 2003).  Mid-IR emission from dust provides an 
independent tracer of star formation.  There are strong
indications that dust-enshrouded star formation is more common
at $z \sim 1$ (e.g. Bell \etal\ 2005; Le Floc'h \etal\ 2005).  
Bright and red galaxies with emission line ratios consistent with star
formation are probably quite dust-reddened; we ask
whether very red objects dominate the SFR,
and whether we find IR-luminous objects
whose emission lines are mostly obscured.



\begin{figure}[ht]
\begin{center}
\includegraphics[width=3.5truein]{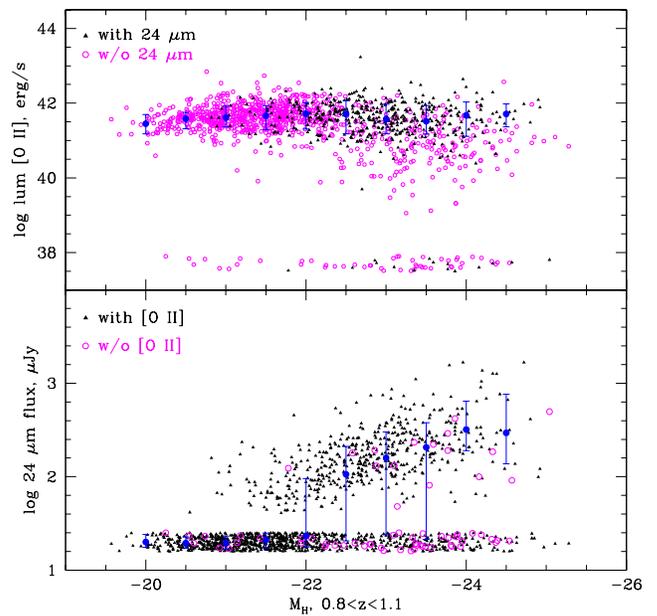}
\caption{
The \oii\ luminosity and 24 micron flux as function of $M_H$
for galaxies with $0.8<z<1.1$.  
Non-detections in either \oii\ or 24 micron are plotted around 
log \oii\ $=37.7$ or log $\stwofour=1.3$, with a random component 
added to spread out the symbols.
For each of $L(\oii)$ and \stwofour, black triangles are galaxies 
detected in the other indicator and magenta open circles are not detected 
in the other.  The bright galaxies with low \oii\ have red colors.
Large blue circles show the median and $\pm 34\%$ for blue galaxies.
}
\label{fig-magoiis24}
\end{center}
\end{figure}

We consider the galaxies within the DEEP2+MIPS region at $0.8<z<1.1$,
where \oii\ is always within the DEEP2 wavelength range.  
Figure \ref{fig-magoiis24} plots the star formation probes \oii\ 
and \stwofour\ as functions of rest $M_H$.  In each panel, the galaxies 
detected/undetected in the {\it other} indicator are shown as black 
triangles or magenta circles respectively.

There is a strong trend of \stwofour\ with magnitude: where $>50\%$ of
the galaxies are detected, the slope of 
the median points in \stwofour\ is $-0.29 \pm 0.05$ dex/$H$ mag.
The medians are unaffected by the exact fluxes of undetected galaxies,
but this slope remains tentative.
The \oii\ luminosity varies much less, by $-0.02$ dex/$H$ mag.
Red galaxies are generally lower in \oii\ than blue.
The 24 $\mu$m flux detects the brightest and reddest of blue galaxies, 
and does not detect many blue galaxies even though they have quite high \oii.
The $\stwofour/\oii$ ratio is a strong function of magnitude, and of color.
This ratio is a probe of extinction, similar to the IR/UV ratio 
(e.g. Wang \& Heckman 1996; Heckman \etal\ 1998; Bell 2003).  
The slope of median $\stwofour/\oii$ is 
$-0.27$ dex/$H$ mag.  But most of this is due to the slopes
of $\oii/\ha$ and extinction, derived earlier; converting
\oii\ to extinction-corrected \ha, the implied slope of 
log $\stwofour/\ha_{extcorr}$ is $-0.04$ dex/$H$ mag -- slightly
weaker than the local measurement by Moustakas \etal\ (2006), with the
caveat on our \stwofour\ slope.  The $\stwofour/\ha_{extcorr}$
slope suggests that extinction is a bit stronger in bright galaxies than the 
Balmer decrement implies, which is found in starburst and IR-luminous 
galaxies (Afonso \etal\ 2003; Cardiel \etal\ 2003; Choi \etal\ 2006).
However, it is also possible that the 24 $\mu$m flux is a
function of dust content and depressed in lower-luminosity, bluer 
systems (Bell 2003).

Of 557 galaxies with $0.8<z<1.1$ and log $\stwofour>1.6$, only 19
were not detected in \oii:
13 are true non-detections while 6 are real \oii\ lines with 
failures of line fitting, so
only 2\% of 24 $\mu$m sources do not show \oii.  Thus there is
not a dominant population of highly obscured galaxies that are 
undetectable in line emission at $z\sim 1$.  Such galaxies could 
still exist if they are extincted so that DEEP2 fails to obtain a 
redshift, or extincted below the DEEP2 magnitude limit.  The former 
cannot be common since most DEEP2 redshift failures
are faint {\it blue} galaxies at high-$z$ (Willmer \etal\ 2006).  
The 24 $\mu$m sources at $z \sim 1$ are not
dominated by obscured red galaxies; most
are luminous intermediate-color galaxies around $U-B=0.8-1.1$
(see also Bell \etal\ 2005; Melbourne, Koo \&
Le Floc'h 2006).  Optically red 24 $\mu$m sources, some of which
may be AGN, are much fewer in number.

\section{Conclusions}

The behavior of emission line ratios is strongly dependent
on galaxy color.  Where possible, star formation studies should 
treat blue and red galaxies separately.  
For the vast majority of blue galaxies,
line ratios follow well-defined relations with magnitude
and are consistent with star formation.  The ratios 
$\ha/\hb$ and $\oii/\hb$ have slopes on $M_H$ of
-0.040 and +0.094 dex/$H$ mag, which may be used to 
calibrate extinction and SFR.  The ratio $\oii_{obs}/\ha_{extcorr}$
has an inferred slope on $M_H$ of +0.23 dex/$H$ mag, which affects
measurements of SFR as a function of mass that use \oii.
Only 1-2\% of blue galaxies have 
ratios indicating AGN activity.  Red galaxies with emission 
have larger scatter in their line ratios and at least half 
show signatures of AGN emission (see also Yan \etal\ 2006).

Mid-IR Spitzer/MIPS 24 $\mu$m fluxes are very valuable as a 
probe of star formation that is unextincted by dust, especially
in high-SFR galaxies.  
The increase of 24 $\mu$m flux with $H$ luminosity is slightly greater 
than the slope inferred for extinction-corrected \ha.
However, 24 $\mu$m does not always detect blue
galaxies with quite high \oii\ luminosity, and we cannot rule
out that 24 $\mu$m strength depends on metallicity and/or dust content.
Perhaps surprisingly, \oii\ detects 98\% of IR-luminous galaxies
at $z \sim 1$; there are few candidates for completely obscured
star formation, and the bright 24 $\mu$m sources are dominated by 
intermediate-color galaxies, not very red ones.  
The different selections and sensitivities to dust of \oii\ 
and 24 $\mu$m samples suggest that neither alone provides a 
complete picture of star formation at $z \sim 1$.  


\acknowledgments

BJW has been supported by grant NSF AST-0242860 to S. Veilleux.  We
acknowledge the cultural role of the summit of Mauna Kea within the
indigenous Hawaiian community and are grateful to have been able to
observe from this mountain.  WIRC observations were obtained at the
Hale Telescope, Palomar Observatory as part of a continuing
collaboration between Caltech,
NASA/JPL, and Cornell University.  This work is based in part on
observations made with the Spitzer Space Telescope and partial support for
this work was provided by NASA to CP through the Spitzer Space Telescope
Fellowship Program, operated by and funded through the Jet Propulsion
Laboratory, California Institute of Technology under a contract with
NASA.  Support for this work was provided by NASA through contract
1255094 issued by JPL/Caltech.






\end{document}